\documentclass[english]{article}
\usepackage[T1]{fontenc}
\usepackage{babel}
\usepackage{graphics}   
\makeatletter

\providecommand{\LyX}{L\kern-.1667em\lower.25em\hbox{Y}\kern-.125emX\@}

\makeatother
\begin{document}

\title{The Physical Basis of Integrable Spin Models}

\author{Indrani Bose,
\\
Department of Physics
,
Bose Institute
,\\
93/1, A.P.C. Road
,
Calcutta-700009, India}
\maketitle

\begin{abstract}
Integrable models are often constructed with real systems in mind.
The exact solvability of the models leads to results which are unambiguous
and provide the correct physical picture. In this review, we discuss
the physical basis of some integrable spin models and their relevance
in the study of real systems. The emphasis in the review is on physical
understanding rather than on the mathematical aspects of integrability.
\end{abstract}

\section{Introduction}

The study of integrable models constitutes an important area of theoretical
physics. Integrable models in condensed matter physics describe interacting
many particle systems. The most prominent examples are interacting
spin and electron systems which include several real materials of
interest. Integrable models, because of their exact solvability, provide
a complete and unambiguous understanding of the variety of phenomena
exhibited by real systems. Integrability in the quantum case implies
the existence of N conserved quantities where N is the number of degrees
of freedom of the system. The corresponding operators including the
Hamiltonian commute with each other. More specifically, integrable
models are also described as exactly-solvable since the ground state
energy and the excitation spectrum of the models can be determined
exactly. Historically, the first example of the exact solvability
of a many body problem was that of a spin\( -\frac{1}{2} \) quantum
spin chain \cite{1}. The technique used to solve the eigenvalue problem
is now known as the Bethe Ansatz (BA) named after Hans Bethe who formulated
it. The demonstration of integrability, namely, the existence of N
commuting operators can be made in the more general mathematical framework
of the Quantum Inverse Scattering Method (QISM) \cite{2}. The BA
has been used extensively to obtain exact results for several quantum
models in one dimension (1d). Examples include the Fermi and Bose
gas models in which particles on a line interact through delta function
potentials \cite{3}, the Hubbard model \cite{4}, 1d plasma which
crystallises as a Wigner solid \cite{5}, the Lai-Sutherland model
which includes the Hubbard model and a dilute magnetic model as special
cases \cite{6}, the Kondo model \cite{7}, the single impurity Anderson
model \cite{8}, the supersymmetric \( t-J \) model (\( J=2t \)
) etc \cite{9}. The BA method has further been applied to derive
exact results for classical lattice statistical models in 2d.

The BA denotes a particular form for the many-particle wave function.
In a 1d system with pairwise interactions, a two particle scattering
conserves the momenta individually due to the energy and momentum
conservation constraints peculiar to 1d. Hence the scattering particles
can either retain their original momenta or exchange them. In the
case of two particles (\( N=2 \) ), the wave function has the form

\begin{equation}
\label{1}
\psi (x_{1},x_{2})=A_{12}e^{i(k_{1}x_{1}+k_{2}x_{2})}+A_{21}e^{i(k_{2}x_{1}+k_{1}x_{2})}
\end{equation}
 where \( x_{1} \) , \( x_{2} \) denote the locations of the two
particles and \( k_{1} \) , \( k_{2} \) are the momentum variables.
The wave function can alternatively be written as

\begin{equation}
\label{2}
\psi (x_{1},x_{2})=e^{i(k_{1}x_{1}+k_{2}x_{2})}+e^{i\theta _{12}}e^{i(k_{2}x_{1}+k_{1}x_{2})}
\end{equation}
where \( \theta _{12} \) is the scattering phase shift. The BA generalises
the wave function (Eq.(1)) to the general case of N particles and
is given by 

\begin{equation}
\label{3}
\psi =\sum _{P}A(P)e^{i\sum ^{N}_{j=1}k_{Pj}x_{j}},x_{1}<x_{2}<.....x_{N}
\end{equation}
The sum over P is a sum over all permutations of \( 1,...,N \). The
amplitude \( A(P) \) is factorisable. Each \( A(P) \) is a product
of factors \( e^{\theta _{ij}} \) 's corresponding to each exchange
of \( k_{i} \) 's required to go from the ordering \( 1,...,N \)
to the ordering P. An overall sign factor may arise depending on the
parity of the permutation. The unknown variables \( \theta _{ij} \) 's and
\( k_{i} \) 's are obtained as solutions of coupled nonlinear equations.
The factorisability condition is at the heart of the exact solvability
of the eigenvalue problem. In the more general QISM approach, the
so-called Yang-Baxter equation provides the condition for factorization
of a multi-particle scattering matrix in terms of two-particle scattering
matrices.

The traditional BA (Eq.(3)) is known as the Coordinate Bethe Ansatz
(CBA). Over the years, the BA method has been generalised in different
ways. The nested BA technique \cite{3,10} has been applied to study
a system of particles with internal degrees of freedom. The state
of a system of electrons is specified in terms of both the spatial
positions as well as the spin indices of the electrons. The Asymptotic
Bethe Ansatz \cite{11} deals with a class of models in which the
interaction between a pair of particles falls off as the inverse square
of the distance between the particles. The Thermodynamic Bethe Ansatz
method \cite{12} is used to calculate thermodynamic quantities and
is a finite temperature extension of the BA method. The Algebraic
Bethe Ansatz (ABA) \cite{13} has been developed in the powerful mathematical
framework of the QISM. The ABA and CBA are equivalent in the sense
that both lead to the same results for the energy eigenvalues. The
CBA, however, does not provide knowledge of the correlation functions
as the structure of the wave function is not sufficiently explicitly
known. The QISM allows the calculation of the correlation functions
in some cases \cite{14}. The mathematical formalism is also much
more systematic and general. One can further establish the existence
of an infinite number (\( N\rightarrow \infty  \)) of mutually commuting
operators. The QISM moreover provides a prescription for the construction
of integrable models. In this review, we will not discuss the mathematical
aspects of integrable models for which a good number of reviews already
exist \cite{2,15,16,17}.  We focus on the physical basis of some
integrable spin models in condensed matter physics and the useful
physical insights derived from the solution of these models. The review
is not meant to be exhaustive and should be supplemented by the references
quoted at the end.

\section{Spin models in 1d}

The interest in 1d spin models arises from the fact that there are
several real magnetic materials which can be described by such models.
The spins interact via the Heisenberg exchange interaction and in
many compounds the exchange interaction within a chain of spins is
much stronger than that between chains. Thus the compounds effectively
behave as linear chain systems. The most general exchange interaction
Hamiltonian describing a chain of spins in which only nearest-neighbour
(n.n.) spins interact is given by

\begin{equation}
\label{4}
H_{XYZ}=\sum ^{N}_{i=1}\left[ J_{x}S^{x}_{i}S^{x}_{i+1}+J_{y}S^{y}_{i}S^{y}_{i+1}+J_{z}S^{z}_{i}S^{z}_{i+1}\right] 
\end{equation}
where \( S^{\alpha }_{i}(\alpha =x,y,z) \) is the spin operator at
the lattice site \( i \), \( N \) is the total number of sites and
\( J_{\alpha } \) denotes the strength of the exchange interaction.
Consider the spins to be of magnitude \( \frac{1}{2} \). The eigenvalue
problems corresponding to the isotropic chain \( (J_{x}=J_{y}=J_{z}=J) \)
and the longitudinally anisotropic chain \( (J_{x}=J_{y}\neq J_{z}) \)
were originally solved using the CBA. Later, the same solutions were
obtained using the formalism of QISM \cite{13,15}. Baxter \cite{18}calculated
the ground state energy of the fully anisotropic model (Eq.(4)) and
Johnson, Krinsky and McCoy \cite{19} found the excitation spectrum.
The results were derived on the basis of a special relationship between
the transfer matrix of the exactly-solved 2d classical lattice statistical
eight vertex model and the fully anisotropic quantum spin Hamiltonian
\( H_{XYZ} \) . Later, the same results were obtained by the ABA
approach of the QISM. The Ising \( (J_{x}=J_{y}=0) \) and the XY
\( (J_{z}=0) \) Hamiltonians are special cases of \( H_{XYZ} \)
.

Consider the isotropic Heisenberg exchange interaction Hamiltonian
in 1d 

\begin{equation}
\label{5}
H=J\sum ^{N}_{i=1}\overrightarrow{S}_{i}.\overrightarrow{S}_{i+1}
\end{equation}
with periodic boundary conditions. The sign of the exchange interaction
determines the favourable alignment of the n.n. spins. \( J>0 \)
corresponds to antiferromagnetic (AFM) exchange interaction due to
which n.n. spins tend to be antiparallel. If \( J<0 \) (equivalently,
replace \( J \) by \( -J \) in Eq.(5) with \( J>0 \) ), the exchange
interaction is ferromagnetic (FM) favouring a parallel alignment of
n.n. spins. One can include a magnetic field term \( -h\sum ^{N}_{i=1}S^{z}_{i} \)
in the Hamiltonian (Eq.(5)), where \( h \) is the strength of the
field. Given a Hamiltonian , the quantities of interest are the ground
state energy and the low-lying excitation spectrum. Knowledge of the
latter enables one to calculate thermodynamic quantities like magnetization,
specific heat and susceptibility at low temperatures. In the case
of the FM Heisenberg Hamiltonian, the exact ground state has a simple
structure. All the spins are parallel, i.e., they align in the same
direction. The lowest excitation is a spin wave or magnon. The excitation
is created by deviating a spin from its ground state arrangement and
letting it propagate. For more than one spin deviation, one has continua
of scattering states as well as bound complexes of magnons. In a bound
complex, the spin deviations preferentially occupy n.n. lattice positions.
The r-magnon bound state energy can be calculated using the BA \cite{1}and
the energy (in units of J) measured w.r.t. the ground state energy
is

\begin{equation}
\label{6}
\epsilon =\frac{1}{r}(1-cosK)
\end{equation}
 where \( K \) is the centre of mass momentum of the \( r \) magnons.
The spin wave excitation energy is obtained for \( r=1 \). The results
can be generalised to the longitudinally anisotropic \( XXZ \) Hamiltonian.
The multimagnon bound states were first detected in the quasi-1d magnetic
system \( CoCl_{2}.2H_{2}O \) \cite{20}. Later improvements made
it possible to observe even \( 14 \) magnon bound states \cite{21}.

In the case of the AFM isotropic Heisenberg Hamiltonian, the ground
state is a singlet and the ground state wave function is a linear
combination of all possible states in which half the spins are up
and the other half down. The AFM ground state can be obtained from
the FM ground state by creating \( r=\frac{N}{2} \) magnons with
momenta \( k_{i} \) and negative energies \( -J(1-cosk_{i}) \).
Remember that the sign of the exchange integral is changed in going
from ferromagnetism to antiferromagnetism. The highest energy state
in the FM case \( (r=\frac{N}{2}) \) becomes the ground state in
the AFM case. The BA equations can be recast in terms of the variables
\( z_{i}\equiv cot(\frac{k_{i}}{2}) \) \cite{22}: 

\begin{equation}
\label{7}
Narctanz_{i}=\pi I_{i}+\sum _{j\neq i}arctan\left( \frac{z_{i-}z_{j}}{2}\right) ,i=1,2,....,r
\end{equation}
 The Bethe quantum numbers \( I_i \)'s 
are integers (half integers) for odd (even) r. For a state specified
by \( \left\{ I_{1},...,I_{r}\right\}  \), the solution \( (z_{1},...,z_{r}) \)
can be obtained from Eq. (7). The energy and the momentum wave number
of the state are given by

\begin{equation}
\label{8}
\frac{E-E_{F}}{J}=-\sum ^{r}_{i=1}\frac{2}{1+z^{2}_{i}}
\end{equation}
 \begin{equation}
\label{9}
k=\pi r-\frac{2\pi }{N}\sum ^{r}_{i=1}I_{i}
\end{equation}
 with \( E_{F}=\frac{JN}{4} \). For the AFM ground state, the Bethe
quantum numbers are given by 

\begin{equation}
\label{10 }
\left\{ I_{i}\right\} =\left\{ -\frac{N}{4}+\frac{1}{2},-\frac{N}{4}+\frac{3}{2},....,\frac{N}{4}-\frac{1}{2}\right\} 
\end{equation}
In the thermodynamic limit \( N\rightarrow \infty  \), the exact
ground state energy has been computed as 

\begin{equation}
\label{11}
E_{g}=NJ(-ln2+\frac{1}{4})
\end{equation}
 The AFM ground state serves as the physical vacuum for the creation
of elementary excitations. These excitations are not the spin-1 magnons
but spin\( -\frac{1}{2} \) spinons \cite{23}. The spinons can be
generated systematically by suitable modifications of the vacuum array
of the BA quantum numbers (Eq.(2)) (for details see \cite{22,23}
). For even N, spinons are always created in pairs, each such pair
originating from the removal of one magnon from the ground state.
Since the spinons are spin\( -\frac{1}{2} \) objects, the lowest
excitations consisting of a pair of spinons are four-fold degenerate
, three triplet \( (S=1) \) and one singlet \( (S=0) \) excitations.
The energy can be written as \( E(k_{1},k_{2})=\epsilon (k_{1})+\epsilon (k_{2}) \)
where the spinon spectrum \( \epsilon (k_{i})=\frac{\pi}{2} \sin{k_{i}} \)
and the total momentum \( k=k_{1}+k_{2} \). At a fixed total momentum
k, one gets a continuum of scattering states. The lower boundary of
the continuum is given by \( \frac{\pi }{2}\left| sink\right|  \)
with one of the \( k_{i}'s=0 \). The upper boundary is obtained for
\( k_{1}=k_{2}=\frac{k}{2} \) and is given by \( \pi \left| sin\frac{k}{2}\right|  \).
Figure 1 gives an example of a two-spinon configuration.

\vspace{0.3cm}
\mbox{ {\resizebox*{!}{2.5cm}{\includegraphics{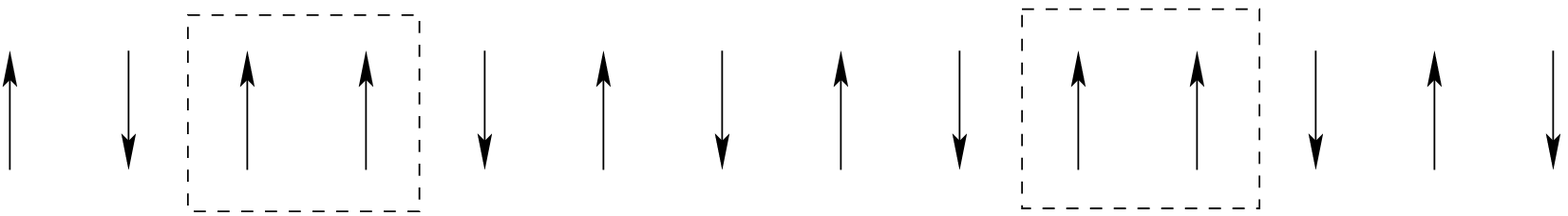}}}}

\vspace{0.3cm}   

\centerline { Figure 1. 
{\it A two-spinon configuration in an AFM chain.}}

\vskip 0.5cm

 The BA results
are obtained in the thermodynamic limit. In this limit, the energies
and the momenta of the spinons just add up, showing that they do not
interact. Since the spinons are excited in pairs, the total spin of
the excited state is an integer. Inelastic neutron scattering study
of the linear chain \( S=\frac{1}{2} \) HAFM compound \( KCuF_{3} \)
has confirmed the existence of unbound spinon pair excitations \cite{24}.
It is to be noted that in the case of a ferromagnet, the low-lying
excitation spectrum consists of a single magnon branch whereas the
AFM spectrum is a two-spinon continuum with well-defined lower and
upper boundaries.

The dynamical properties of a magnetic system are governed by the
time-dependent pair correlation functions or their space-time double
Fourier transforms known as dynamical correlation functions. An important
time-dependent correlation function is

\begin{equation}
\label{12}
G(R,t)=\left\langle \overrightarrow{S}_{R}(t).\overrightarrow{S}_{0}(0)\right\rangle 
\end{equation}
 The corresponding dynamical correlation function is the quantity
measured in inelastic neutron scattering experiments. The differential
scattering cross-section in such an experiment is given by 

\begin{equation}
\label{13}
\frac{d^{2}\sigma }{d\Omega d\omega }\propto S^{\mu \mu }(\overrightarrow{q},\omega )=\frac{1}{N}\sum _{R}e^{i\overrightarrow{q}.\overrightarrow{R}}\int ^{+\infty }_{-\infty }dte^{i\omega t}\left\langle S^{\mu }_{R}(t)S^{\mu }_{0}(0)\right\rangle 
\end{equation}
where \( \overrightarrow{q} \) and \( \omega  \) are the momentum
wave vector and energy of the spin excitation and \( \mu =x,y,z \).
For a particular \( \overrightarrow{q} \) , the peak in \( S^{\mu \mu }(\overrightarrow{q},\omega ) \)
occurs at a value of \( \omega  \) which gives the excitation energy.
At \( T=0 \), 

\begin{equation}
\label{14}
S^{\mu \mu }(\overrightarrow{q},\omega )=\sum _{\lambda }M^{\mu }_{\lambda }\delta (\omega +E_{g}-E_{\lambda })
\end{equation}
 \( E_{g}(E_{\lambda }) \) is the energy of the ground (excited)
state and 

\begin{equation}
\label{15 }
M^{\mu }_{\lambda }=2\pi \left| \left\langle G\right| S^{\mu }(\overrightarrow{q})\left| \lambda \right\rangle \right| ^{2}
\end{equation}
 is the transition rate between the singlet (\( S_{tot}=0 \)) ground
state \( \left| G\right\rangle  \) and the triplet (\( S_{tot}=1 \))
states \( \left| \lambda \right\rangle  \) \cite{25}. The exact
calculation of the dynamical correlation functions in the BA formalism
is not possible. Bougourzi et al \cite{26} have used an alternative
approach, based on the algebraic analysis of the completely integrable
spin chain, and have calculated the exact 2-spinon part of the dynamical
correlation function \( S^{xx}(q,\omega ) \) for the 1d \( S=\frac{1}{2} \)
AFM \( XXZ \) model. In this model, the Ising part of the \( XXZ \)
Hamiltonian provides the dominant interaction. Karbach et al have
\cite{27}calculated the exact 2-spinon part of \( S^{zz}(q,\omega ) \)
for the isotropic Heisenberg Hamiltonian. In both the cases, the size
of the chain is infinite. The exact form of the 2-spinon contribution
to the dynamical correlation function \( S^{xx}(q,\omega ) \) of
the \( S=\frac{1}{2} \) \( XXZ \) HAFM chain is complicated and
is given by

\begin{equation}
\label{16}
S_{(2)}^{xx}(q,\omega )=\frac{\omega _{0}}{8I\omega }\left[ 1+\sqrt{\frac{\omega ^{2}-\chi ^{2}\omega _{0}^{2}}{\omega ^{2}-\omega ^{2}_{0}}}\right] \sum _{c=\pm }\frac{\vartheta ^{2}_{A}(\beta ^{c}_{-})}{\vartheta ^{2}_{d}(\beta ^{c}_{-})}\frac{\left| tan(\frac{q}{2})\right| ^{-c}}{W_{c}}
\end{equation}
where \( I=\frac{JK}{\pi }sinh\frac{\pi K^{\prime }}{K},\chi \equiv \frac{1-k^{\prime }}{1+k^{\prime }} \)and
\( k,k^{\prime }\equiv \sqrt{1-k^{2}} \) are the modulii of the elliptic
integrals \( K\equiv K(k),K^{\prime }\equiv K(k^{\prime }) \). The
anisotropy parameter \( q=-exp(\frac{-\pi K^{\prime }}{K}) \) with
\( -\frac{J_{z}}{J_{x}}=\Delta =(\frac{q+q^{-1}}{2}) \). Also,

\begin{equation}
\label{17}
W_{\pm }=\sqrt{\frac{\omega ^{4}_{0}}{\omega ^{4}}\chi ^{2}-(\frac{T}{\omega ^{2}}\pm cosq)^{2}}
\end{equation}
 \begin{equation}
\label{18}
T=\sqrt{\omega ^{2}-\chi ^{2}\omega ^{2}_{0}}\sqrt{\omega ^{2}-\omega ^{2}_{0}}
\end{equation}
 \begin{equation}
\label{19}
\omega _{0}=\frac{2Isin(q)}{1+\chi }
\end{equation}
 \begin{equation}
\label{20}
\beta ^{c}_{-}(q,\omega )=\frac{1+\chi }{2}F\left[ arcsin(\frac{2I\omega W_{c}}{\chi (1+\chi )\omega ^{2}_{0}}),\chi \right] 
\end{equation}
 (\( F \) is the incomplete elliptic integral)

\begin{equation}
\label{21}
\vartheta ^{2}_{A}(\beta )=exp(-\sum ^{\infty }_{l=1}\frac{e^{\gamma l}}{l}\frac{cosh(2\gamma l)cos(t\gamma l)-1}{sinh(2\gamma l)cosh(\gamma l)})
\end{equation}
 \( \gamma =\frac{\pi K^{\prime} }{K},t\equiv \frac{2\beta }{K^{\prime}} \)
and \( \vartheta _{d}(x) \) is a Neville theta function. The derivation
of \( S_{(2)}^{xx}(q,\omega ) \) involves generating the 2-spinon
states from the spinon vacuum, namely, the AFM ground state, with
the help of spinon creation operators and expressing the spin fluctuation
operator \( S^{\mu }(q) \) in terms of the spinon creation operators.
The 2-spinon part is expected to provide the dominant contribution
to the dynamical correlation function (Eq. (14)). For example, in
the case of the isotropic Heisenberg Hamiltonian, the 2-spinon excitations
account for approximately \( 73\% \) of the total intensity in \( S^{zz}(q,\omega ) \).
The 2-spinon triplet excitations play a significant role in the low-temperature
spin dynamics of quasi-1d AFM compounds like \( KCuF_{3},Cu(C_{6}D_{5}COO)_{2}.3D_{2}O,Cs_{2}CuCl_{4} \)
and \( Cu(C_{4}H_{4}N_{2}(NO_{3})_{2}) \) \cite{24,28}. These excitations
can be probed via inelastic neutron scattering and hence a knowledge
of the exact dynamical correlation function is useful. The 2-spinon
singlet excitations cannot be excited in neutron scattering because
of selection rules (the spinon vacuum \( \left| G\right\rangle  \)
is a singlet and the excited state \( \left| \lambda \right\rangle  \)
in Eq.(15) is a triplet). Linear chain compounds like \( CuGeO_{3} \)
exhibit the spin-Peierls transition \cite{29}. The transition gives
rise to lattice distortion and consequently to a dimerization of the
exchange interaction. Exchange interactions between successive pairs
of spins alternate in strength. There is a tendency for the formation
of dimers (singlets) across the strong bonds. One can construct an
appropriate dynamical correlation function in which the dimer fluctuation
operator (DFO) replaces the spin fluctuation operator \( S^{\mu }(q) \).
The DFO connects the AFM ground state to the 2-spinon singlet and
not to the 2-spinon triplet. 

Two well-known physical realizations of the 1d \( S=\frac{1}{2} \)
Ising-Heisenberg compounds are \( CsCoCl_{3} \) and \( CsCoBr_{3} \).
Several inelastic neutron scattering measurements have been carried
out on these compounds to probe the low-temperature spin dynamics
\cite{30}. In these compounds, the Ising part of the \( XXZ \) Hamiltonian
is significantly dominant so that perturbation calculations around
the Ising limit are feasible. Near the Ising limit, the exact 2-spinon
dynamical correlation function \( S^{xx}(q,\omega ) \) is identical
in the lowest order to the first-order perturbation result of Ishimura
and Shiba (IS) \cite{31}. The IS calculation provides physical insight
on the nature of spinons. The Ising part of the \( XXZ \) Hamiltonian
is the unperturbed Hamiltonian and the \( XY \) part constitutes
the perturbation. The two-fold degenerate N\'{e}el states are the
ground states of the Ising Hamiltonian. These two states serve as
the {}``spinon vacuua''. An excitation is created by flipping a
block of adjacent spins from the spin arrangement in the N\'{e}el
state. For example, in Figure 1, a block of seven spins is flipped
in the N\'{e}el state. The block of overturned spins gives rise to
two parallel spin pairs at its boundary with the unperturbed N\'{e}el
configuration. It is these domain walls or kink solitons which are
the equivalents of spinons. A 2-spinon excited state \( (S^{z}_{tot}=1) \)
is obtained as a linear superposition of states in which an odd number
\( \nu  \) (\( \nu =1,3,5,.... \)) of spins is overturned in the
N\'{e}el configuration. In each such state, both the domain walls
have equal spin orientations with the spins pointing up. The excitation
continuum of two spinons is obtained in first order perturbation theory.
The lineshapes of \( S^{xx}(q,\omega ) \) observed in experiments
are highly asymmetric with a greater concentration of intensity near
the spectral threshold and a tail extending to the upper boundary
of the continuum. The exact 2-spinon part of \( S^{xx}(q,\omega ) \)
has also an asymmetric shape in agreement with experimental data.
The first order perturbation-theoretic result of IS for \( S^{xx}(q,\omega ) \)
fails to reproduce the asymmetry. A second-order perturbation calculation
leads to greater asymmetry in the lineshapes \cite{32}. Furthermore,
in the framework of a first order perturbation theory, the effects
of full anisotropy \( (J_{x}\neq J_{y}\neq J_{z}) \), next-nearest-neighbour
coupling, interchain coupling and exchange mixing have been shown
to give rise to asymmetry in lineshapes \cite{33}. 

Recently, a large number of studies have been carried out on a class
of models in which the interaction between spins falls off as the
inverse-square of the distance between them. A lattice model which
belongs to this class is known as the Haldane-Shastry model \cite{34}
the Hamiltonian of which is given by

\begin{equation}
\label{22}
H=J\sum _{i<j}\frac{P_{ij}}{d(i-j)^{2}}
\end{equation}
 where \( d(l)=(\frac{N}{\pi })\left| sin\frac{\pi l}{N}\right|  \)is
the chord distance between the pair of spins separated by \( l \)
sites on a ring with \( N \) equally spaced spins. \( P_{ij} \)
is the spin exchange operator, \( P_{ij}=(2\overrightarrow{S}_{i}.\overrightarrow{S}_{j}+\frac{1}{2}) \).
The model is exactly solvable and the key results are: the ground
state has a form similar to the fractional quantum Hall ground state,
the ground state is a QSL and the elementary excitations are the spin\( -\frac{1}{2} \)
spinons obeying fractional statistics, the thermodynamics as well
as the various dynamical correlation functions can be calculated exactly.
The latter calculations are possible because of the simple structure
of the eigenspectrum. 

A correct analysis of the BA equations for the \( S=\frac{1}{2} \)
HAFM in 1d gave rise to the concept of spinons which has subsequently
been verified in experiments. Approximate methods like spin wave theory
fail to predict the spinon continuum thus pointing to the importance
of integrable models in providing the correct physical picture. The
existence of spinons in dimension greater than one is a highly debatable
issue. No precise statement can be made due to the lack of exact results
in \( d>1 \). The issue is of considerable significance in connection
with the resonating-valence-bond (RVB) theory of high temperature
superconductivity. In a valence bond (VB) state, pairs of spins are
in singlet spin configurations (a singlet is often termed as a VB).
The RVB state is a coherent linear superposition of VB states. In
1973, Anderson \cite{35} in a classic paper suggested that the ground
state of the \( S=\frac{1}{2} \) HAFM on the frustrated triangular
lattice is a RVB state. The RVB state is a singlet (total spin is
zero) and is often described as a quantum spin liquid (QSL) since
translational as well as rotational symmetries are preserved in the
state. The RVB state is spin disordered and the two-spin correlation
function has an exponential decay as a function of the distance between
the spins. Interest in the RVB state revived after the discovery of
high temperature superconductivity in 1987\cite{36}. The common structural
ingredient of the high\( -T_{C} \) cuprate systems is the copper-oxide
\( (CuO_{2}) \) plane which ideally behaves as a \( S=\frac{1}{2} \)
HAFM defined on a square lattice. It is largely agreed that the ground
state \( (T=0) \) has AFM long range order (LRO). The low-lying excitations
are the conventional \( S=1 \) magnons. In the spinon picture, a
magnon is a pair of confined spinons. The spinons cannot move apart
from each other unlike in 1d. The cuprates exhibit a rich phase diagram
as a function of the dopant concentration. On doping, positively charged
holes are introduced in the \( CuO_{2} \) plane. The holes are mobile
in a background of antiferromagnetically interacting spins. The motion
of holes acts against antiferromagnetism and the AFM LRO is rapidly
destroyed as the concentration of holes increases. The resulting spin
disordered state has been speculated to be a RVB state. In close analogy
with the \( S=\frac{1}{2} \) HAFM chain, the low-lying spin excitations
in the RVB state are pairs of spinons. The spinons are created by
breaking a VB. The spinons are not confined as in the case of an ordered
ground state but separate via a rearrangement of the VBs. The spinons
have spin \( \frac{1}{2} \) and charge \( 0 \). The charge excitations
in a RVB state are known as holons with charge \( +e \) and spin
\( 0 \). Holons are created on doping the RVB state, i.e., replacing
electrons by holes. Spinons and holons are best decribed as topological
excitations in a QSL. The key feature of the doped RVB state is that
of spin-charge separation, i.e., the spin and charge excitations are
decoupled entities. Spin-charge separation can be rigorously demonstrated
in the case of interacting electron systems in 1d known by the general
name of Luttinger Liquids (LLs). The Hubbard model in 1d is the most
well-known example of a LL. The model is integrable and the BA results
for the excitation spectrum confirm that the spinons and the holons
are the elementary excitations \cite{36,37}.

Coming back to the RVB state, there has been an intensive search for
spin models in 2d with RVB states as exact ground states. Recent calculations
show that there is AFM LRO in the ground state of the \( S=\frac{1}{2} \)
HAFM on the triangular lattice, contrary to Anderson's original conjecture
\cite{38}. Frustrated spin models with n.n. as well as non-n.n. exchange
interactions have been constructed for which the RVB states are the
exact ground states in certain parameter regimes \cite{39}. These
are short-ranged RVB states with the VBs forming between n.n. spin
pairs. The spinon excitation spectrum in this case is gapped. A model
which captures the low energy dynamics in the RVB scenario is the
Quantum Dimer Model (QDM)\cite{40}. The Hamiltonian of the model
defined on a square lattice is given by 

\vspace{0.3cm}
\begin{equation}
\label{23}
\mbox{ {\resizebox*{!}{1.5cm}{\includegraphics{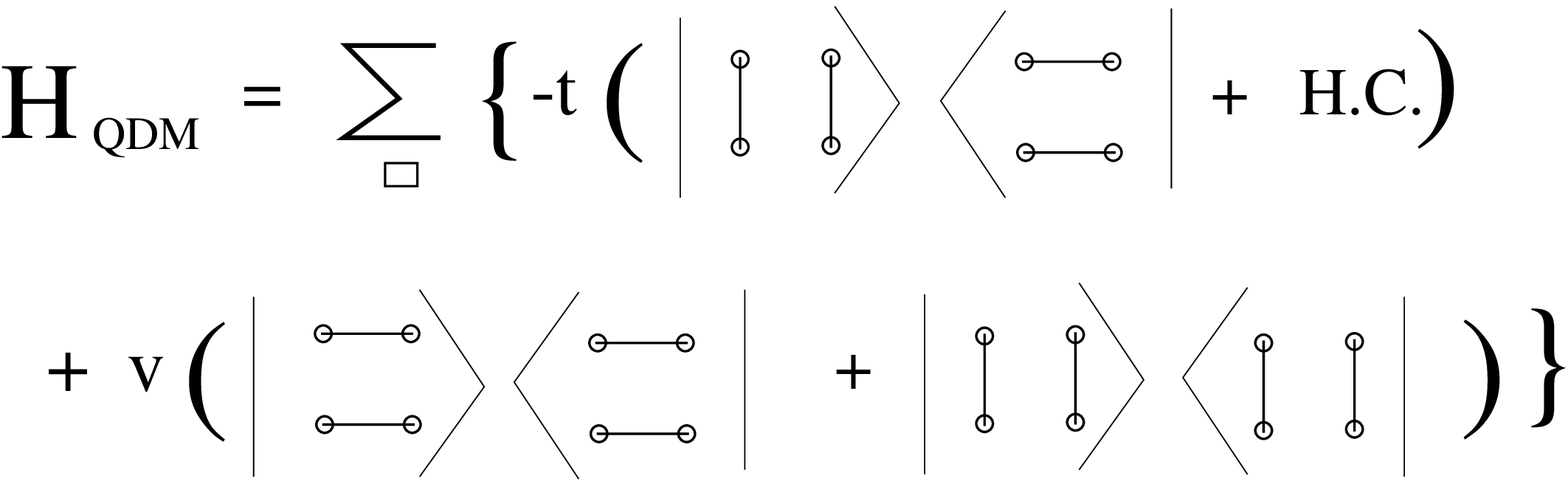}}}}
\end{equation}
\vspace{0.3cm}   

where the solid lines represent dimers (VBs) and the sum runs over
all the plaquettes of the lattice. The first term of the Hamiltonian
is the kinetic part representing the flipping of a pair of parallel
dimers on the two bonds of a plaquette to the other possible orientation,
i.e., from horizontal to vertical and vice versa. The second term
counts the number of flippable pairs of dimers in any dimer configuration
and is analogous to the potential term of the Hamiltonian. The ground
state of the QDM on the square lattice is not, however, a QSL except
at the special point \( t=V \). Moessner and Sondhi \cite{41} have
studied the QDM on the triangular lattice and shown that, in contrast
to the square lattice case , the ground state is a RVB state with
deconfined, gapped spinons in a finite range of parameters. Recently,
some microscopic models of 2d magnets have been proposed \cite{42}
the low-lying excitations of which are of three types: spinons, holons
and {}``vortex-like'' excitations with no spin and charge, dubbed
as visons. Some of these models are related to the QDM. Two integrable
models \cite{42,43} have been constructed which share common topological
features with the microscopic models in 2d and have applications in
fault-tolerant quantum computation. The models ,however, cannot resolve
the issue of spinons in 2d as quantum numbers like the total \( S^{z} \)
are not conserved in these models. The search for microscopic models
in 2d, with spinons as elementary excitations, acquires particular
significance in the light of recent experimental evidence of the spinon
continuum in the 2d frustrated quantum antiferromagnet \( Cs_{2}CuCl_{4} \)
\cite{44}. The ground state of this compound is expected to be a
QSL with spinons and not magnons as elementary excitations. Exactly
solvable models in 2d are needed for a clear understanding of the
origin of the experimentally observed spinon continuum.

Real materials are often anisotropic in character. The anisotropy
may be present in the exchange interaction Hamiltonian itself or there
may be additional terms in the Hamiltonian corresponding to different
types of anisotropy. A well-known anisotropic interaction, present
in many AFM materials, is the Dzyaloshinskii-Moriya (DM) interaction
with the general form

\begin{equation}
\label{24}
H_{DM}=\overrightarrow{D}.(\overrightarrow{S}_{i}\times \overrightarrow{S}_{j})\textrm{ }
\end{equation}
Moriya \cite{45} provided the microscopic basis of the DM interaction
by extending Anderson's superexchange theory to include the spin-orbit
interaction. The DM coupling acts to cant the spins because the coupling
energy is minimised when the two spins are perpendicular to each other.
Some examples of materials with DM interaction include the quasi-2d
compound \( Cs_{2}CuCl_{4} \) \cite{44}, the \( CuO_{2} \) planes
of the undoped cuprate system \( La_{2}CuO_{4} \) \cite{46}, the
quasi-1d compound Cu-Benzoate \cite{47} etc. The DM canting of spins
is responsible for the small ferromagnetic moment of the \( CuO_{2} \)
planes even though the dominant in-plane exchange interaction is AFM
in nature. Alcaraz and Wreszinski \cite{48} have shown that the \( XXZ \)
quantum Heisenberg chain (both FM and AFM) with DM interaction is
equivalent to the \( XXZ \) Hamiltonian with modified boundary conditions
and anisotropy parameter \( \frac{J_{z}}{J_{x}} \). The DM interaction
is assumed to be of the form

\begin{equation}
\label{25}
H_{DM}(\Delta )=-\frac{\Delta }{2}\sum ^{N}_{i=1}(\sigma ^{x}_{i}\sigma ^{y}_{i+1}-\sigma ^{y}_{i}\sigma ^{x}_{i+1})
\end{equation}
,i.e., the vector \( \overrightarrow{D} \) in Eq.(23) is in the z-direction.
The new anisotropy parameter is \( \frac{\delta }{\sqrt{1+\Delta ^{2}}} \)
where \( \delta  \) is the anisotropy parameter of the original \( XXZ \)
Hamiltonian. With changed boundary conditions, the model is still
BA solvable. In fact, in the thermodynamic limit \( (N\rightarrow \infty ) \),
the boundary conditions do not affect the critical behaviour. Thus,
the Hamiltonian, which includes both the \( XXZ \) Hamiltonian and
the DM interaction, has the same critical properties and the phase
diagram as the \( XXZ \) Hamiltonian with the anisotropy parameter
\( \frac{\delta }{\sqrt{1+\Delta ^{2}}} \). 

We next turn our attention to spin\( -S \) (\( S>\frac{1}{2} \))
quantum spin chains. The spin\( -S \) Heisenberg exchange interaction
Hamiltonian in 1d is not integrable. A family of Heisenberg-like models
has been constructed for \( S=1,\frac{3}{2},2,\frac{5}{2},... \)
etc. for which the spin\( -S \) quantum Hamiltonian is given by 

\begin{equation}
\label{26}
H_{s}=\sum _{i}Q(\overrightarrow{S}_{i}.\overrightarrow{S}_{i+1})
\end{equation}
 where \( Q(x) \) is a polynomial of degree \( 2S \) \cite{49}.
With this generalization, the spin\( -S \) quantum spin chains are
integrable. The integrable models, however, do not distinguish between
half-odd integer and integer spins. In both the cases, the integrable
models have gapless excitation spectrum. For half-odd integer AFM
Heisenberg spin chains (with only the bilinear exchange interaction
term), the Lieb-Schultz-Mattis (LSM) theorem \cite{50} states that
the excitation spectrum is gapless. The theorem cannot be proved for
AFM integer spin chains. Haldane in 1983 pointed out the difference
between the half-odd integer and integer AFM Heisenberg spin chains
and made the conjecture that integer spin chains have a gap in the
excitation spectrum \cite{51}. Integer spin quantum antiferromagnets
in 1d have been widely studied analytically, numerically and experimentally
and Haldane's conjecture has turned out to be true. There are several
examples of quasi-1d \( S=1 \) AFM materials which exhibit the Haldane
gap. Some of the most widely studied materials are \( CsNiCl_{3},Ni(C_{2}H_{8}N_{2})_{2}NO_{2}(ClO_{4}) \)
(NENP), \( Y_{2}BaNiO_{5} \) etc. Recently, experimental evidence
of a \( S=2 \) antiferromagnet which exhibits the Haldane gap has
been obtained. In this compound the manganese ions form effective
\( S=2 \) spins and are coupled in a quasi-1d chain \cite{52}. Integrable
models of integer spin chains do not reproduce the Haldane gap but
are of considerable interest since they provide exact information
about the phase diagram of generalised integer spin models. Consider
the generalised Hamiltonian for an AFM \( S=1 \) chain:

\begin{equation}
\label{27}
H=\sum _{i}\left[ cos\theta (\overrightarrow{S}_{i}.\overrightarrow{S}_{i+1})+sin\theta (\overrightarrow{S}_{i}.\overrightarrow{S}_{i+1})^{2}\right] 
\end{equation}
with \( \theta  \) varying between \( 0 \) and \( 2\pi  \). The
biquadratic term has been found to be relevant in some real integer-spin
materials. There are two gapped phases: the Haldane phase for \( -\frac{\pi }{4}<\theta <\frac{\pi }{4} \)and
a dimerised phase for \( -\frac{3\pi }{4}<\theta <-\frac{\pi }{4} \)
\cite{53}. At \( \theta =-\frac{\pi }{4} \), the model is integrable
and the gap vanishes to zero. This point separates the two gapped
phases, Haldane and dimerised, which have different symmetry properties.
Thus a quantum phase transition occurs at \( \vartheta =-\frac{\pi }{4} \)
from the Haldane to the dimerised phase. The integrable model provides
exact location of the transition point. The point \( \theta =\frac{\pi }{4} \)corresponds
to the Hamiltonian which is a sum over permutation operators and is
again exactly solvable. The Haldane phase includes the isotropic Heisenberg
chain \( (\theta =0) \) and the Affleck-Kennedy-Lieb-Tasaki (AKLT)
Hamiltonian \( (tan\theta _{VBS}=\frac{1}{3}) \) \cite{54}. The
latter model is not integrable but the ground state is known exactly.
The ground state is described as a valence bond solid (VBS) state
in which a VB (singlet) covers every link of the chain. Since the
gap does not become zero for \( 0\leq \theta \leq \theta _{VBS} \),
there is no phase transition in going from one limiting Hamiltonian
to the other. Thus the isotropic Heisenberg and AKLT chains are in
the same phase.

The doped cuprate systems exhibit a variety of novel phenomena in
their insulating, metallic and superconducting phases. A full understanding
of these phenomena is as yet lacking. There is currently a strong
research interest in doped spin systems. The idea is to look for simpler
spin systems in which the consequences of doping can be studied in
a less ambiguous manner. The spin\( -1 \) HG nickelate compound \( Y_{2}BaNiO_{5} \)
can be doped with holes on replacing the off-chain \( Y^{3+} \) ions
by \( Ca^{2+} \) ions. Inelastic neutron scattering (INS) measurements
on the doped compound provide evidence for the appearance of new states
in the HG \cite{55}. The structure factor \( S(q) \), obtained by
integrating the dynamical correlation function \( S(q,\omega ) \)
over \( \omega  \), acquires an incommensurate, double-peaked form
in the doped state \cite{56}. Frahm et al \cite{57} have constructed
an integrable model describing a doped spin\( -1 \) chain. In the
undoped limit, the spectrum is gapless and so the HG of the integer
spin system is not reproduced. It is, however, possible to reintroduce
a gap in the continuum limit where a field-theoretical description
of the model is possible. The model has limited relevance in explaining
the physical features of the doped nickelate compound. Another interesting
study relates to the appearance of magnetization plateaus in the doped
\( S=1 \) integrable model \cite{58}. The location of the plateaus
depends on the concentration of holes. Experimental evidence of this
novel phenomenon has not been obtained so far. 

An electron in a solid, localised around an atomic site, has three
degrees of freedom charge, spin and orbital. The orbital degree of
freedom is relevant to several transition metal oxides which include
the cuprate and manganite systems. The latter compounds on doping
exhibit the phenomenon of colossal magnetoresistance in which there
is a huge change in electrical resistivity on the application of a
magnetic field. The manganites like the cuprates have a rich phase
diagram as a function of the dopant concentration \cite{59}. We now
give a specific example of the orbital degree of freedom. The \( Mn^{3+} \)
ion in the manganite compound \( LaMnO_{3} \) has four electrons
in the outermost \( 3d \) energy level. The electrostatic field of
the neighbouring oxygen ions splits the 3d energy level into two sublevels,
\( t_{2g} \) and \( e_{g} \). Three of the four electrons occupy
the three \( t_{2g} \) orbitals \( d_{xy},d_{yz,}d_{zx} \) and the
fourth electron goes to the \( e_{g} \)-sublevel containing the two
orbitals \( d_{x^{2}-y^{2}} \) and \( d_{3z^{2}-r^{2}} \). The fourth
electron thus has an orbital degree of freedom as it has two possible
choices for occupying an orbital. The four electrons have the same
spin orientation to minimise the electrostatic repulsion energy according
to the Hund's rule. The total spin is thus \( S=2 \). The orbital
degree of freedom is described by the pseudospin \( \overrightarrow{T} \)
such that \( T_{z}=\frac{1}{2}(-\frac{1}{2}) \) when the \( d_{x^{2}-y^{2}} \)
\( (d_{3z^{2}-r^{2}}) \) orbital is occupied. The three components
of the pseudospin satisfy commutation relations similar to those of
the spin components. The \( e_{g} \) doublet is further split into
two hyperfine energy levels due to the well-known Jahn-Teller (JT)
effect. In concentrated systems, the JT effect can lead to orbital
ordering below an ordering temperature. In the antiferromagnetically
ordered N\'{e}el state, the spins are alternately up and down. Similarly,
in the case of antiferroorbital ordering, the occupied orbitals alternate
in type at successive sites of the lattice. The orbital degree of
freedom is frozen as a result. Apart from the JT mechanism of orbital
ordering, there is an exchange mechanism which may lead to orbital
order. The exchange mechanism is a generalisation of the usual superexchange
to the case of orbital degeneracy. Starting from the degenerate Hubbard
model, in which there are two degenerate orbitals at each site, one
can derive the following generalised exchange Hamiltonian \cite{60}:

\begin{equation}
\label{28}
H=\sum _{ij}\left\{ J_{1}\overrightarrow{S}_{i}.\overrightarrow{S}_{j}+J_{2}\overrightarrow{T}_{i}.\overrightarrow{T}_{j}+J_{3}(\overrightarrow{S}_{i}.\overrightarrow{S}_{j})(\overrightarrow{T}_{i}.\overrightarrow{T}_{j})\right\} 
\end{equation}
 Consider the case \( J_{1}=J_{2}=J \). For \( J_{3}=0 \), two independent
Heisenberg-like Hamiltonians are obtained which are BA solvable. At
the Kolezhuk-Mikeska point, \( \frac{J_{3}}{J}=\frac{4}{3} \), the
ground state is exactly known \cite{61}. The point \( \frac{J_{3}}{J}=4 \)
is integrable and there are three gapless excitation modes. The compounds
\( Na_{2}Ti_{2}Sb_{2}O \) and \( NaV_{2}O_{5} \) are examples of
materials in 1d with coupled spin and orbital degrees of freedom \cite{62}.
These systems have been described by anisotropic versions of the Hamiltonian
in Eq.(28) but without adequate agreement with experiments. The elementary
excitations in the orbital sector are the orbital waves or {}``orbitons''.
An excitation of this type is created in the orbitally ordered state
by changing the occupied orbital at a site and letting the defect
propagate in the solid. The excitations are analogous to the spin
waves or magnons in a magnetically ordered solid. Experimental evidence
of orbital waves has recently been obtained in the manganite compound
\( LaMnO_{3} \) through Raman scattering measurements \cite{63,64}.
As discussed before, integrable spin models provide important links
between theory and experiments. A similar scenario in the case of
systems with coupled spin and orbital degrees of freedom is yet to
develop.

\section{Ladder models}

\mbox{ {\resizebox*{!}{2.5cm}{\includegraphics{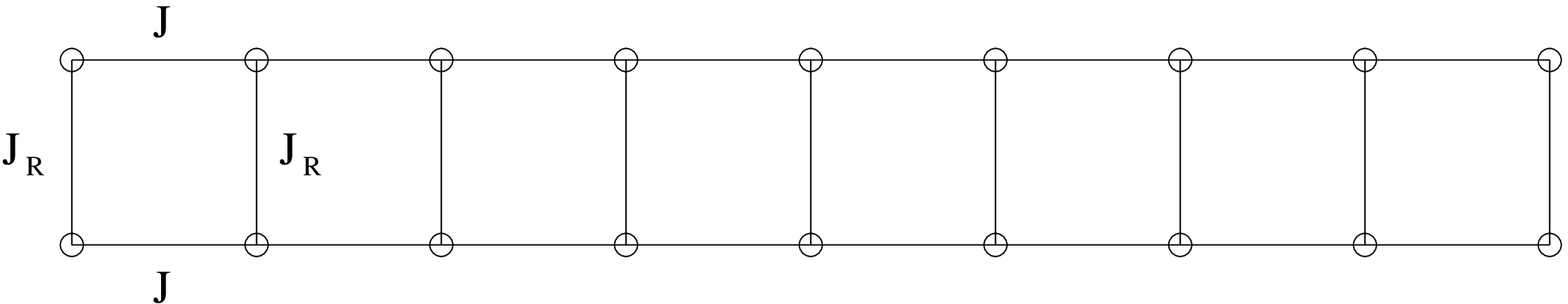}}}}

\vspace{0.3cm}   

\centerline { Figure 2. 
{\it  A two-chain ladder. The rung and intra-chain n.n. exchange
interactions}} \centerline { 
{\it are of strength \( J_{R} \) and \( J \) respectively.
 }}

\vskip 0.5cm

The simplest ladder model consists of two chains coupled by rungs
(Figure 2). In general, the ladder may consist of n chains coupled
by rungs. In the spin ladder model, each site of the ladder is occupied
by a spin (in general of magnitude \( \frac{1}{2} \)) and the spins
interact via the Heisenberg AFM exchange interaction. In the doped
spin ladder model, some of the sites are empty, i.e., occupied by
holes. The holes can move in the background of interacting spins.
There are two major reasons for the considerable research interest
in ladders. Powerful techniques like the BA and bosonization are available
for the study of 1d many body systems whereas practically very few
rigorous results are known for 2d systems. Ladders provide a bridge
between 1d and 2d physics and are ideally suited to study how the
electronic and magnetic properties change as one goes from a single
chain to the square lattice. The unconventional properties of the
\( CuO_{2} \) planes of the cuprate systems are the main reason for
the significant interest in 2d many body systems. Many of these properties
are ascribed to strong correlation effects. Ladders are simpler systems
in which some of the issues associated with strong correlation can
be addressed in a more rigorous manner. The second motivation for
the study of ladder systems is that several such systems have been
discovered in the recent past. In the following, we describe in brief
some of the major physical properties of ladders. There are two exhaustive
reviews on ladders which provide more detailed information \cite{65,66}. 

Consider a two-chain spin ladder described by the AFM Heisenberg exchange
interaction Hamiltonian

\begin{equation}
\label{29}
H=\sum _{\left\langle ij\right\rangle }J_{ij}\overrightarrow{S}_{i}.\overrightarrow{S}_{j}
\end{equation}
 The n.n. intra-chain and the rung exchange interactions are of strength
\( J \) and \( J_{R} \) respectively. When \( J_{R}=0 \), one obtains
two decoupled AFM spin chains for which the excitation spectrum is
known to be gapless. For all \( \frac{J_{R}}{J}>0 \), a gap (the
so-called spin gap (SG)) opens up in the spin excitation spectrum.
The result is easy to understand in the simple limit in which the
exchange coupling \( J_{R} \) along the rungs is much stronger than
the coupling \( J \) along the chains. The intra-chain coupling may
thus be treated as perturbation. When \( J=0 \), the exact ground
state consists of singlets along the rungs. The ground state energy
is \( -\frac{3J_{R}N}{4} \), where \( N \) is the number of rungs
in the ladder. The ground state has total spin \( S=0 \). In first
order perturbation theory, the correction to the ground state energy
is zero. A \( S=1 \) excitation may be created by promoting one of
the rung singlets to a \( S=1 \) triplet. The weak coupling along
the chains gives rise to a propagating \( S=1 \) magnon. In first
order perturbation theory, the dispersion relation is 

\begin{equation}
\label{30}
\omega (k)=J_{R}+Jcosk
\end{equation}
 where \( k \) is the momentum wave vector. The SG defined as the
minimum excitation energy is given by 

\begin{equation}
\label{31}
\Delta _{SG}=\omega (\pi )\simeq (J_{R}-J)
\end{equation}
 The two-spin correlations decay exponentially along the chains showing
that the ground state is a quantum spin liquid (QSL). The magnons
can further form bound states. Experimental evidence of two-magnon
bound states has been obtained in the \( S=\frac{1}{2} \)two-chain
ladder compound \( Ca_{14-x}La_{x}Cu_{24}O_{41} \) \( (x=5 \) and
\( 4) \) \cite{67}. The family of compounds \( Sr_{n-1}Cu_{n+1}O_{2n} \)
consists of planes of weakly-coupled ladders of \( \frac{n+1}{2} \)chains
\cite{68}. For \( n=3 \) and \( 5 \), respectively, one gets the
two-chain and three-chain ladder compounds \( SrCu_{2}O_{3} \) and
\( Sr_{2}Cu_{3}O_{5} \) respectively. For the first compound, experimental
evidence of the SG has been obtained. The latter compound has properties
similar to those of the 1d Heisenberg AFM chain \cite{69}. A recent
example of a spin ladder belonging to the organic family of materials
is the compound \( (C_{5}H_{12}N)_{2}CuBr_{4} \), a ladder system
with strong rung coupling \( (\frac{J_{R}}{J}\simeq 3.5) \) \cite{70}.
The phase diagram of the AFM spin ladder in the presence of an external
magnetic field is particularly interesting. In the absence of the
magnetic field and at \( T=0 \), the ground state is a QSL with a
gap in the excitation spectrum. At a field \( H_{c_{1}} \), there
is a transition to a gapless Luttinger Liquid (LL) phase \( (g\mu _{B}H_{c_{1}}=\Delta _{SG} \),
the spin gap, \( \mu _{B} \) is the Bohr magneton and \( g \) the
Land\'{e} splitting factor). There is another transition at an upper
critical field \( H_{c_{2}} \) to a fully polarised FM state. Both
\( H_{c_{1}} \) and \( H_{c_{2}} \) are quantum critical points.
The quantum phase transition from one ground state to another is brought
about by changing the magnetic field. At small temperatures, the behaviour
of the system is determined by the crossover between two types of
critical behaviour: quantum critical behaviour at \( T=0 \) and classical
critical behaviour at \( T\neq 0 \). Quantum effects are persistent
in the crossover region at small finite temperatures and such effects
can be probed experimentally. In the case of the ladder system \( (C_{5}H_{12}N)_{2}CuBr_{4} \),
the magnetization data obtained experimentally exhibit universal scaling
behaviour in the vicinity of the critical fields \( H_{c_{1}} \)
and \( H_{c_{2}} \). In the gapless regime \( H_{c_{1}}<H<H_{c_{2}} \),
the ladder model can be mapped onto an \( XXZ \) chain the thermodynamic
properties of which can be calculated exactly by the BA. The theoretically
computed magnetization \( M \) versus magnetic field \( h \) curve
is in excellent agreement with the experimental data. Organic spin
ladders provide ideal testing grounds for the theories of quantum
phase transitions. For inorganic spin ladder systems, the value of
\( H_{c_{1}} \) is too high to be experimentally accessible.

Bose and Gayen \cite{71} have studied a frustrated two-chain spin
model with diagonal couplings. The intrachain and diagonal spin-spin
interactions are of equal strength \( J \). It is easy to show that
for \( J_{R}\geq 2J \), the exact ground state consists of singlets
(dimers) along the rungs with the energy \( E_{g}=-\frac{3J_{R}N}{4} \)
where \( N \) is the number of rungs. Xian \cite{72} later pointed
out that as long as \( \frac{_{J_{R}}}{J}>\left( \frac{_{J_{R}}}{J}\right) _{c}\simeq 1.401 \),
the rung dimer state is the exact ground state. At \( \frac{J_{R}}{J}=\left( \frac{J_{R}}{J}\right) _{c} \)
, there is a first order transition from the rung dimer state to the
Haldane phase of the \( S=1 \) chain. Kolezhuk and Mikeska \cite{73}
have constructed a class of generalised \( S=\frac{1}{2} \) two-chain
ladder models for which the ground state can be determined exactly.
The Hamiltonian \( H \) is a sum over plaquette Hamiltonians and
each such Hamiltonian contains various two-spin as well as four-spin
interaction terms. They have further introduced a toy model which
has a rich phase diagram in which the phase boundaries can be determined
exactly. 

The standard spin ladder models with bilinear exchange are not integrable.
For integrability, multispin interaction terms have to be included
in the Hamiltonian. Some integrable ladder models have already been
constructed \cite{74}. We discuss one particular model proposed by
Wang \cite{75}. The Hamiltonian is given by

\begin{eqnarray}
H & = & \frac{J_{1}}{4}\sum ^{N}_{i=1}\left[ \overrightarrow{\sigma }_{j}.\overrightarrow{\sigma }_{j+1}+\overrightarrow{\tau }_{j}.\overrightarrow{\tau }_{j+1}\right] +\frac{J_{2}}{2}\sum ^{N}_{j=1}\overrightarrow{\sigma }_{j}.\overrightarrow{\tau }_{j}\nonumber \\
 & + & \frac{U_{1}}{4}\sum ^{N}_{j=1}\left( \overrightarrow{\sigma }_{j}.\overrightarrow{\sigma }_{j+1}\right) \left( \overrightarrow{\tau }_{j}.\overrightarrow{\tau }_{j+1}\right) +\frac{U_{2}}{4}\sum ^{N}_{j=1}\left( \overrightarrow{\sigma }_{j}.\overrightarrow{\tau }_{j}\right) \left( \overrightarrow{\sigma }_{j+1}.\overrightarrow{\tau }_{j+1}\right) \label{32} 
\end{eqnarray}
where \( \overrightarrow{\sigma }_{j} \) and \( \overrightarrow{\tau }_{j} \)
are the Pauli matrices associated with the site \( j \) of the upper
and lower chains respectively. \( N \) is the total number of rungs
in the system. The ordinary spin ladder Hamiltonian is obtained from
Eq. (32) when the four spin terms are absent, i.e., \( U_{1}=U_{2}=0 \).
For general parameters \( J_{1},J_{2},U_{1} \) and \( U_{2} \),
the model is non-integrable. The integrable cases correspond to \( U_{1}=J_{1},U_{2}=0 \)
or \( U_{1}=J_{1},U_{2}=-\frac{J_{1}}{2} \). Without loss of generality
one can put \( J_{1}=U_{1}=1,J_{2}=J \) and \( U_{2}=U \). For \( U=0 \),
the Hamiltonian (32) reduces to

\begin{equation}
\label{33}
H=\frac{1}{4}\sum ^{N}_{j=1}(1+\overrightarrow{\sigma }_{j}.\overrightarrow{\sigma }_{j+1})(1+\overrightarrow{\tau }_{j}.\overrightarrow{\tau }_{j+1})+\frac{J}{2}\sum ^{N}_{j=1}(\overrightarrow{\sigma }_{j}.\overrightarrow{\tau }_{j}-1)+\frac{1}{2}(J-\frac{1}{2})N
\end{equation}
Three quantum phases are possible. For \( J>J^{c}_{+}=2 \), the system
exists in the rung dimerised phase. The ground state is a product
of singlet rungs. The SG is given by \( \Delta _{SG}=2(J-2) \). For
\( J^{c}_{+}>J>J^{c}_{-} \), a gapless phase is obtained with three
branches of gapless excitations. \( J^{c}_{+} \) is the quantum critical
point at which a QPT from the dimerised phase to the gapless phase
occurs. In the vicinity of the quantum critical point, the susceptibility
and the specific heat can be calculated using the thermodynamic BA.
From the low-temperature expansion of the thermodynamic BA equation,
one obtains 

\begin{equation}
\label{34}
C\sim T^{\frac{1}{2}},\chi \sim T^{-\frac{1}{2}}
\end{equation}
 which are typical of quantum critical behaviour. In the presence
of an external magnetic field \( h \), the magnetic field can be
tuned to drive a QPT at the quantum critical point \( h_{c}=2(J-2) \)
from the gapless phase to a gapped phase. The third quantum phase
\( (h=0) \) is obtained for \( J<J^{c}_{-}=-\frac{\pi }{4\sqrt{3}}+\frac{ln3}{4} \).
This is a gapless phase with two branches of gapless excitations.
For \( U=-\frac{1}{2} \), a similar phase diagram is obtained. Note
that the ladder model may equivalently be considered as a spin-orbital
model with \( \overrightarrow{\sigma } \) and \( \overrightarrow{\tau } \)
representing the spin and the pseudospin.

Doped ladder models are toy models of strongly correlated systems
\cite{65}. In these systems, the double occupancy of a site by two
electrons, one with spin up and the other with spin down, is prohibited
due to strong coulomb correlations. In a doped spin system, there
is a competition between two processes: hole delocalization and exchange
energy minimization. A hole moving in an antiferromagnetically ordered
spin background, say, the N\'{e}el state, gives rise to parallel spin
pairs which raise the exchange interaction energy of the system. The
questions of interest are: whether a coherent motion of the holes
is possible, whether two holes can form a bound state, the development
of superconducting (SC) correlations, the possibility of phase separation
of holes etc. Some of these issues are of significant relevance in
the context of doped cuprate systems in which charge transport occurs
through the motion of holes \cite{76}. In the SC phase, the holes
form bound pairs with possibly d-wave symmetry. Several proposals
have been made so far on the origin of hole binding but there is as
yet no general consensus on the actual binding mechanism. The doped
cuprate systems exist in a `pseudogap' phase before the SC phase is
entered. In fact, some cuprate systems also exhibit SG. As already
mentioned, the doped two-chain ladder systems are characterised by
a SG. The issue of how the gap evolves on doping is of significant
interest. The possibility of binding of hole pairs in a two-chain
ladder system was first pointed out by Dagotto et al \cite{77}. In
this case, the binding mechanism is not controversial and can be understood
in a simple physical picture. Again, consider the case \( J_{R}\gg J \),
i.e., a ladder with dominant exchange interactions along the rungs.
In the ground state, the rungs are mostly in singlet spin configurations.
On the introduction of a single hole, a singlet spin pair is broken
and the corresponding exchange interaction energy is lost. When two
holes are present, they prefer to be on the same rung to minimise
the loss in the exchange interaction energy. The holes thus form a
bound pair. In the more general case, detailed energy considerations
show that the two holes tend to be close to each other and effectively
form a bound pair. For more than two holes, several calculations suggest
that considerable SC pairing correlations develop in the system on
doping. True superconductivity can be obtained only in the bulk limit.
Theoretical predictions motivated the search for ladder compounds
which can be doped with holes. Much excitement was created in 1996
when the ladder compound \( Sr_{14-x}Ca_{x}Cu_{24}O_{41} \) was found
to become SC under pressure at \( x=13.6 \) \cite{78}. The transition
temperature \( T_{c} \) is \( \sim 12K \) at a pressure of \( 3GPa \).
As in the case of cuprate systems, bound pairs of holes are responsible
for charge transport in the SC phase. Experimental results on doped
ladder compounds point out strong analogies between the doped ladder
and cuprate systems \cite{65}. 

The strongly correlated doped ladder system is described by the t-J
Hamiltonian

\begin{equation}
\label{35}
H_{t-J}=-\sum _{\left\langle ij\right\rangle ,\sigma }t_{ij}(\widetilde{C}_{i\sigma }^{+}\widetilde{C}_{j\sigma }+H.C.)+\sum _{\left\langle ij\right\rangle }J_{ij}(\overrightarrow{S}_{i}.\overrightarrow{S}_{j}-\frac{1}{4}n_{i}n_{j})
\end{equation}
The \( \widetilde{C}_{i\sigma }^{+} \) and \( \widetilde{C}_{i\sigma } \)
are the electron creation and annihilation operators which act in
the reduced Hilbert space (no double occupancy of sites), 

\begin{eqnarray}
\widetilde{C}_{i\sigma }^{+} & = & C^{+}_{i\sigma }(1-n_{i-\sigma })\nonumber \\
\widetilde{C}_{i\sigma } & = & C_{i\sigma }(1-n_{i-\sigma })\label{36} 
\end{eqnarray}
where \( \sigma  \) is the spin index and \( n_{i} \), \( n_{j} \)
are the occupation numbers of the \( ith \) and \( jth \) sites
respectively. The first term in Eq.(35) describes the motion of holes
with hopping integrals \( t_{R} \) and \( t \) for motion along
the rung and chain respectively. In the standard \( t-J \) ladder
model, \( i \) and \( j \) are n.n. sites. The second term contains
the usual AFM Heisenberg exchange interaction Hamiltonian. The \( t-J \)
model thus describes the motion of holes in a background of antiferromagnetically
interacting spins. A large number of studies have been carried out
on \( t-J \) ladder models. These are reviewed in Refs. \cite{65,66}.
We describe briefly some of the major results. The SG of the undoped
ladder changes discontinuously on doping. Remember that the SG is
the difference in energies of the lowest triplet excitation and the
ground state. In the doped state, there are two distinct triplet excitations.
One triplet excitation is that of the undoped ladder obtained by exciting
a rung singlet to a rung triplet. A new type of triplet excitation
is possible when at least two holes are present. On the introduction
of two holes in two rung singlets, a pair of free spin\( -\frac{1}{2} \)'s
is obtained which combines to give rise to a singlet \( (S=0) \)
or a triplet \( (S=1) \) state. The triplet configuration of the
two free spins corresponds to the second type of triplet excitation.
The SG of this new excitation is unrelated to the SG of the magnon
excitation. The true SG is the one which has the lowest value in a
particular parameter regime.

The low energy modes of a ladder system are characterised by their
spin. Singlet and triplet excitations correspond to charge and spin
modes respectively. In each sector, the hole may further be in a bonding
or antibonding state with opposite parities. We consider only the
even parity sector to which the lowest energy excitations belong.
In both the \( S=0 \) and \( S=1 \) sectors, an excitation continuum
with well-defined boundaries is present. The \( S=0 \) and the \( S=1 \)
continua are degnerate in energy. A bound state branch with \( S=0 \)
splits off below the continuum the lowest energy of which corresponds
to the c.o.m. momentum wave vector \( K=0 \) \cite{79,80}. Thus
the two-hole ground state is in the singlet sector and corresponds
to a bound state of two holes with \( K=0 \). The bound state has
\( d- \)wave type symmetry. Within the bound state branch, excitations
with energy infinitesimally close to the ground state are possible.
These excitations are the charge excitations since the total spin
is still zero and the charge excitation spectrum is gapless. The lowest
spin excitations in a wide parameter regime are between the \( S=0 \)
ground state and the lowest energy state in the \( S=1 \) continuum
\cite{81}. The continuum does not exist in the undoped ladder and
so the SG evolves discontinuously on doping in this parameter regime.
A suggestion has, however, been made that the lowest triplet excitation
is a bound state of a magnon with a pair of holes \cite{82}. In summary,
the two-chain ladder model has the feature that the charge excitation
is gapless but the spin excitation has a gap. This is the Luther-Emery
phase and is different from the LL phase in which both the spin and
charge excitations are gapless.

Bose and Gayen have derived several exact, analytical results for
the ground state energy and the low-lying excitation spectrum of the
frustrated \( t-J \) ladder doped with one and two holes. The undoped
ladder model has already been described. In the doped case, the hopping
integral has the value \( t_{R} \) for hole motion along the rungs
and the intra-chain and diagonal hopping integrals are of equal strength
\( t \). The latter assumption is crucial for the exact solvability
of the eigenvalue problem in the one and two hole sectors. Though
the model differs from the standard \( t-J \) ladder model (the diagonal
couplings are missing in the latter), the spin and charge excitation
spectra exhibit similar features. In particular, the dispersion relation
of the two-hole bound state branch is obtained exactly and the exact ground
state is shown to be a  bound state of two holes with \( K=0 \) and \( d- \)wave
type symmetry. The ladder exists in the Luther-Emery phase. There
is no spin charge separation. as in the case of a LL. In the exact
hole eigenstates, the hole is always accompanied by a free spin\( -\frac{1}{2} \).
The hole-hole correlation function can also be calculated exactly.
When \( J_{R}\gg J \), the holes of a bound pair are predominantly
on the same rung. For lower values of \( J_{R} \), the holes prefer
to be on n.n. rungs so that energy gain through the delocalization
of a hole along the rung is possible.

The t-J ladder model constructed by Bose and Gayen is not integrable.
Frahm and Kundu \cite{84} have constructed a \( t-J \) ladder model
which is integrable. The Hamiltonian is given by

\begin{equation}
\label{37}
H=\sum _{a}H^{(a)}_{t-J}+H_{int}+H_{rung}-\mu \widehat{n}
\end{equation}
The two chains of the ladder are labelled by \( a=1,2 \) and \( \mu  \)
is the chemical potential coupling to the number of electrons in the
system. \( H^{(a)}_{t-J} \) is the \( t-J \) Hamiltonian (Eq.(35))
for a chain plus the terms \( n^{(a)}_{j}+n^{(a)}_{j+1} \)where \( n^{(a)}_{j} \)is
the total number of electrons on site \( j \). 

\begin{equation}
\label{38}
H_{int}=-\sum _{j}\left[ H^{(1)}_{t-J}\right] _{jj+1}\left[ H_{t-J}^{(2)}\right] _{jj+1}
\end{equation}
 \( H_{rung} \) includes the \( t-J \) Hamiltonian (Eq. (35)) corresponding
to a rung and a Coulomb interaction term \( V\sum _{j}n^{(1)}_{j}n^{(2)}_{j} \).
The possible basis states of a rung are the following. When no hole
is present, a rung can be in a singlet or a triplet spin configuration.
When a single hole is present, the rung is in a bonding \( \left( \left| \sigma _{+}\right\rangle \right)  \)
or antibonding \( \left( \left| \sigma _{-}\right\rangle \right)  \)
state with \( \left| \sigma _{\pm }\right\rangle \equiv \frac{1}{\sqrt{2}}\left( \left| \sigma 0\right\rangle \pm \left| 0\sigma \right\rangle \right)  \)
and \( \sigma =\uparrow  \) or \( \downarrow  \). The rung can further
be occupied by two holes. Frahm and Kundu have studied the phase diagram
of the ladder model at low temperatures and in the strong coupling
regime \( J_{R}\gg 1,V\gg \mu +\left| t_{R}\right|  \) near half-filling.
In this regime, the triplet states are unfavourable. By excluding
the triplet states and choosing \( J=2t=2 \), the Hamiltonian \( H \)
(Eq.(37)) can be rewritten as 

\begin{equation}
\label{39}
H=-\sum _{j}\Pi_{jj+1}-\sum ^{5}_{l=1}A_{l}N_{l}+const.
\end{equation}
 where \( N_{l},l=1,2(3,4) \) is the number of bonding (antibonding)
single hole rung states with spin \( \uparrow ,\downarrow  \) and
\( N_{5} \) is the number of empty rungs. If \( L \) is the total
number of rungs in the ladder, the remaining \( N_{0}=L-\sum _{l}N_{l} \)
rungs are in singlet spin configurations. The permutation operator
\( \Pi_{jk} \) interchanges the states on rungs \( j \) and \( k \).
If both the rungs are singly occupied by a hole, an additional minus
sign is obtained on interchanging the rung states. The potentials
\( A_{l} \)'s are: 

\begin{equation}
\label{40}
A_{1}=A_{2}\equiv \mu _{+}=t_{R}-\mu +V
\end{equation}
 \begin{equation}
\label{41}
A_{3}=A_{4}\equiv \mu _{-}=-t_{R-}\mu +V
\end{equation}
 \begin{equation}
\label{42}
A_{5}\equiv \widetilde{V}=-2\mu +V
\end{equation}
 The nature of the ground state and the low-lying excitation spectrum
depends on the relative strengths of the potentials \( A_{l} \)'s.
The Hamiltonian (39) is BA solvable. The phase diagram \( V \) vs.
the hole concentration \( n_{h} \) has been computed for \( \mu _{+}=\mu _{-} \),i.e.,
\( t_{R}=0 \). For large repulsive \( V \), the ground state can
be described as a Fermi sea of single hole states \( \left| \sigma _{\pm }\right\rangle  \)
propagating in a background of rung dimer states \( \left| s\right\rangle  \).
The double-hole rung states \( \left| d\right\rangle  \) are energetically
favourable for sufficiently strong attractive rung interactions. In
the intermediate region, both types of hole rung states are present.
In the frustrated \( t-J \) ladder model studied by Bose and Gayen
\cite{83}, the exact two-hole ground state is a linear combination
of single-hole and double-hole rung states propagating in a background
of rung dimer states. The single-hole rung states are the bonding
states. 

In a remarkable paper, Lin et al \cite{85} have considered the problem
of electrons hopping on a two-chain ladder. The interaction between
the electrons is sufficiently weak and finite-ranged. At half filling,
a perturbative renormalization group (RG) calculation shows that the
model scales onto the Gross-Neveu (GN) model which is integrable and
has \( SO(8) \) symmetry. At half filling, the two-chain ladder is
in the Mott insulating phase with d-wave pairing correlations. The
insulating phase is further a QSL. The integrability has been utilised
to determine the exact energies and quantum numbers of all the low
energy excitations which constitute the degenerate \( SO(8) \) multiplets.
The lowest-lying excitations can be divided into three octets all
with a non-zero gap (mass gap) m. Each excitation has a dispersion
\( \epsilon _{1}(q)=\sqrt{m^{2}+q^{2}} \) where \( q \) is the momentum
variable measured w.r.t. the minimum energy value. One octet consists
of two-particle excitations: two charge \( \pm 2e \) Cooper pairs
around zero momentum, a triplet of \( S=1 \) magnons around momentum
\( (\pi ,\pi ) \) and three neutral \( S=0 \) particle-hole pair
excitations. \( SO(8) \) transformations rotate the components of
the vector multiplet into one another unifying the excitations in
the process. The \( SO(5) \) subgroup which rotates only the first
five components of the vector is the symmetry proposed by Zhang \cite{86} to
unify antiferromagnetism and superconductivity in the cuprates. The
vector octet is related by a triality symmetry to two other octets
with mass gap \( m \). The \( 16 \) particles of these two octets
have the features of quasi-electrons and quasi-holes. Above the 24
states with mass gap \( m \), there are other higher-lying {}``bound''
states with mass gap \( \sqrt{3}m \). Finally, the continuum of scattering
states occurs above the energy \( 2m \). Lin et al has further studied
the effects of doping a small concentration of holes into the Mott
insulating phase. In this limit, the effect of doping can be incorporated
in the GN model by adding a term \( -\mu Q \) to the Hamiltonian,
\( \mu  \) being the chemical potential and \( Q \) the total charge.
Integrability of the GN model is not lost as \( Q \) is a global
\( SO(8) \) generator. Doping is possible only for \( 2\mu >m \)
when Cooper pairs enter the system. The doped ladder exists in the
Luther-Emery phase, whereas in the half-filled insulating limit both
the spin and charge excitations are gapped. In the doped phase, the
Cooper pairs can transport charge and quasi-long-range \( d- \)wave
SC pairing correlations develop in the system. The other features
of the standard \( t-J \) ladder model, e.g., the discontinuous
evolution of the SG on doping is reproduced. The lowest triplet excitation
is a bound state of a \( S=1 \) magnon with a Cooper pair. As mentioned
before, a similar
result has been obtained numerically in the case of the standard \( t-J \)
ladder \cite{82}. The triplet excitation belongs to the family of
\( 28 \) excitations with mass gap \( \sqrt{3}m. \) If \( x \)
denotes the dopant concentration, then the SG jumps from \( \Delta _{S}(x=0)=m \)
to \( \Delta _{S}(x=0^{+})=(\sqrt{3}-1)m \) upon doping. The integrability
of the weakly-interacting two-chain ladder model has yielded a plethora
of exact results which illustrate the rich physics associated with
undoped and doped ladders.

\section{Concluding Remarks}

Integrable models have a dual utility. They serve as testing grounds
for approximate methods and techniques. Also, they are often models
of real systems and provide rigorous information about the physical
properties of such systems. Integrable models are sometimes more general
than what are required to describe real systems. In such cases, an
integrable model corresponds to an exactly solvable point in the general
phase diagram. The point may be a quantum critical point at which
transition from one quantum phase to another occurs or the integrable
model may be in the same phase as a more realistic model. In the latter
case, the physical properties of the two models are similar. In this
review, we have discussed the physical basis of some integrable spin
models with special focus on the relevance of the models to real systems.
The Heisenberg spin chain is probably the best example of the essential
role played by exact solvability in correctly interpreting the experimental
data. The concept of spinons owes its origin to the exact analysis
of the BA equations. The theoretical prediction motivated the search
for real spin systems in which experimental confirmation could be
made. In this review, examples are also given of systems for which
the links between integrable models and experimental results are not
well established. A major portion of the review is devoted to physical
systems which exhibit rich phenomena, like the systems with both spin
and orbital degrees of freedom and undoped and doped spin ladder systems,
where the need for integrable systems is particularly strong. These
systems exhibit a variety of novel phenomena a proper understanding
of which should be based on rigorous theory. Two-dimensional spin
systems with QSL ground states have been specially mentioned to explain
the recent interest in constructing integrable models of such systems.
The review is meant to be an elementary introduction to the genesis
and usefulness of integrable models vis-\`{a}-vis physical spin systems.
Future challenges are also highlighted to motivate further research
on integrable models.

There are some AFM spin models which are not integrable but for which the ground states and in some cases the low-lying excited states are known exactly. The most prominent amongst these are the Majumdar-Ghosh (MG) chain \cite{81} and the AKLT \cite{54} model respectively. The MG Hamiltonian is defined in 1d for spins of magnitude $\frac{1}{2}$. The Hamiltonian includes both n.n. as well as n.n.n. interactions. The strength of the latter is half that of the former. The exact ground state is doubly degenerate and the states consist of singlets along alternate links of the lattice. The excitation spectrum is not exactly known and has been calculated on the basis of a variational wave function \cite{88}. Generalizations of the MG model to 2d with exactly-known ground states are possible \cite{39,89,90,91}. The Shastry-Sutherland model \cite{89} is of much current interest due to the recent discovery of the compound $SrCu_{2}(BO_{3})_{2}$ which is well-described by the model \cite{92}. Some of these models including the AKLT model have been reviewed in the references \cite{93,94,95,96} from which more information about the models can be obtained. These models incorporate physical features of real systems and provide valuable insight on the magnetic properties of low-dimensional quantum spin systems. The models supplement integrable models in obtaining exact information and provide motivation for the construction of integrable generalisations.

\end{document}